\documentclass[conference]{IEEEtran}%,compsoc
\pdfoutput=1 

\ifCLASSOPTIONcompsoc
  \usepackage[nocompress]{cite}
\else
  \usepackage{cite}
\fi

\usepackage[T1,hyphens]{url}
\usepackage{graphics}
\usepackage{tikz}
\usepackage{multirow}
\usepackage{amsmath}
\usepackage{booktabs}
\usepackage{color}
\usepackage{xstring}
\usepackage{amsfonts}
\usepackage{threeparttable}
\usepackage{amsthm}

\begin{document}

\title{Diving Deep into Clickbaits: Who Use Them to What Extents in Which Topics with What Effects?}

\author{
\IEEEauthorblockN{Md Main Uddin Rony{$~^{\S}$}, Naeemul Hassan{$~^{\S}$}, Mohammad Yousuf{$~^{\natural}$}}
\IEEEauthorblockA{{$~^{\S}$}Department of Computer and Information Science, {$~^{\natural}$}Gaylord College of Journalism and Mass Communication\\
{$~^{\S}$}The University of Mississippi, {$~^{\natural}$}The University of Oklahoma}
}

\maketitle

%Reminder
\newcommand{\reminder}[1]{\textbf{\textit{Reminder: }\color{blue}{#1}}}
\newcommand{\updatethis}[1]{\textbf{\color{red}{#1}}}

%Dataset

\newcommand{\dataset}[1]{%
    \IfEqCase{#1}{%
        {1}{Headlines}%
        {2}{Media Corpus}%
        % you can add more cases here as desired
    }[\PackageError{tree}{Undefined option to tree: #1}{}]%
}%

%Style
\newcommand*{\rom}[1]{\expandafter\@slowromancap\romannumeral #1@}

%Method
\newcommand{\googlewv}{Google\_word2vec}
\newcommand{\clkbtwv}{Skip-Gram$_{sw}$}
\newcommand{\glove}{GLOVE}

%Misc
\newtheorem{example}{Example}

\begin{abstract}
The use of alluring headlines (clickbait) to tempt the readers has become a growing practice nowadays. For the sake of existence in the highly competitive media industry, most of the on-line media including the mainstream ones, have started following this practice. Although the wide-spread practice of clickbait makes the reader's reliability on media vulnerable, a large scale analysis to reveal this fact is still absent. In this paper, we analyze $1.67$ million Facebook posts created by $153$ media organizations to understand the extent of clickbait practice, its impact and user engagement by using our own developed clickbait detection model. The model uses distributed sub-word embeddings learned from a large corpus.  The accuracy of the model is 98.3\%. Powered with this model, we further study the distribution of topics in clickbait and non-clickbait contents.
\end{abstract}
\section{introduction}
\label{sec-introduction}

% +What is a clickbait?\\
The term \emph{clickbait} refers to a form of web content that employs writing formulas and linguistic techniques in headlines to trick readers into clicking links \cite{palau2016reference,chakraborty2016stop}, but does not deliver on promises~\footnote{https://www.wired.com/2015/12/psychology-of-clickbait/}. Media scholars and pundits consistently show clickbait content in a bad light, but the industry based on this type of content has been rapidly growing and reaching more and more people across the world ~\cite{mark2016adele,chris2016clickbait}. \emph{Taboola}, one of the key providers of clickbait content, claims \footnote{https://www.taboola.com/press-release/taboola-crosses-one-billion-user-mark-second-only-facebook-world’s-largest-discovery} to have doubled its monthly reach from $500$ million unique users to $1$ billion in a single year from March 2015. The growth of clickbait industry appears to have clear impact on the media ecosystem, as many traditional media organizations have  started to use clickbait techniques to attract readers and generate revenue. However, media analysts suggest that news media risk losing readers' trust and depleting brand value by using clickbait techniques that may boost advertising revenue only temporarily. According to a study performed by Facebook ~\footnote{https://www.nytimes.com/2014/08/26/business/media/facebook-takes-steps-against-click-bait-articles.html}, $80\%$ users ``preferred headlines that helped them decide if they wanted to read the full article before they had to click through''. ~\cite{joshua2016ashley} shows that clickbait headlines lead to negative reactions among media users.

Compared to the reach of clickbait content and its impact on the online media ecosystem, the amount of research done on this topic is very small. No large scale study has been conducted to examine the extent to which different types of media use clickbait techniques. Little is known about the extent to which clickbait headlines contribute to user engagement on social networking platforms -- major distributors of web content. This study seeks to fill this gap by examining uses of clickbait techniques in headlines by mainstream and unreliable media organizations on the social network. Some of the questions we answer in this paper are-- (i) to what extent, mainstream and unreliable media organizations use clickbait? (ii) does the topic distribution of the contents vary in clickbaity contents? (iii) which type of headlines -- clickbait or non-clickbait —- generates more user engagement (e.g., shares, comments, reactions)?

We first create a set of supervised clickbait classification models to identify clickbait headlines. Instead of following the traditional bag-of-words and hand-crafted feature set approaches, we take a more recent deep learning path that does not require feature engineering. Specifically, we use distributed subword embedding technique ~\cite{bojanowski2016enriching,joulin2016bag} to transform the words in the corpus to $300$ dimensional embeddings. These embeddings are used to map sentences to a vector space over which a softmax function is applied as a classifier. Our best performing model achieves $98.3\%$ accuracy on a labeled dataset. We use this model to analyze a larger dataset which is a collection of approximately $1.67$ million Facebook posts created during 2014--2016 by $68$ mainstream media and $85$ unreliable media organizations. In addition to identifying the clickbait headlines in the corpus, we also use the embeddings to measure the distance between the headline and the first paragraph, known as intro, of a news article. We use a word co-occurrence based topic model that learns topics by modeling word-word co-occurrences patterns (e.g., bi-terms) to understand the distribution of topics in the clickbait and non-clickbait contents of each media. Finally, using the data on Facebook reactions, comments, and shares, we analyzed the role clickbaits play in user engagement and information spread. The main contributions of this paper are--

\textbullet We collect a large data corpus of $1.67$ million Facebook posts by over $150$ U.S. based media organizations. Details of the corpus is explained in Section ~\ref{sec-dataset}. We make the corpus available to use for research purpose ~\footnote{URL will be added after acceptance}.

\textbullet We prepare distributed subword based embeddings for the words present in the corpus. In Section ~\ref{sec-clickbaitDetection}, we provide a comparison between these word embeddings and the \emph{word2vec} ~\cite{mikolov2013distributed,mikolov2013efficient} embeddings created from Google News dataset with respect to clickbait detection. We plan to make these embeddings publicly available upon acceptance of the paper.

\textbullet We perform detailed analysis of the clickbait practice in the social network from multiple perspectives. Section ~\ref{sec-clickbaitPractice} presents qualitative, quantitative and impact analysis of clickbait and non-clickbait contents.   
\section{Dataset}
\label{sec-dataset}
We use two datasets in this paper.
%Basic statistics of the datasets are presented in Table~\ref{table-datasets}.
Below, we provide description of the datasets and explain the collection process.

\textbf{\dataset{1}}: This dataset is curated by Chakraborty et al.~\cite{chakraborty2016stop}. It contains $32,000$ headlines of news articles which appeared in `WikiNews', `New York Times', `The Guardian', `The Hindu', `BuzzFeed', `Upworthy', `ViralNova', `Thatscoop', `Scoopwhoop', and `ViralStories'.~\footnote{https://github.com/bhargaviparanjape/clickbait/tree/master/dataset} Each of these headlines is manually labeled either as a clickbait or a non-clickbait by at least three volunteers.
%Details of the collection process can be found in ~\cite{chakraborty2016stop}.
There are $15,999$ clickbait headlines and $16,001$ non-clickbait headlines in this dataset. We used this labeled dataset to develop an automatic clickbait classification model (details in Section ~\ref{sec-clickbaitDetection}). An earlier version of this dataset was used in~\cite{chakraborty2016stop,anand2016we}. It had $15,000$ manually labeled headlines with an even distribution of $7,500$ clickbait and $7,500$ non-clickbait headlines.

\textbf{\dataset{2}}: For large scale analysis,  using Facebook Graph API~\footnote{https://developers.facebook.com/docs/graph-api}, we accumulated all the Facebook posts created by a set of mainstream and unreliable media within January $1^{st}$, 2014 -- December $31^{st}$, 2016. The mainstream set consists of the $25$ most circulated print media ~\footnote{https://en.wikipedia.org/wiki/List\_of\_newspapers\_in\_the\_United\_States} and the $43$ most-watched broadcast media ~\footnote{www.indiewire.com/2016/12/cnn-fox-news-msnbc-nbc-ratings-2016-winners-losers-1201762864/} (according to Nielson rating~\cite{nielson2016fcc}). The unreliable set is a collection of $85$ conspiracy, clickbait, satire and junk science based media organizations. The category of each unreliable media is cross-checked by two sources ~\cite{info2016list,melissa2016list}. Figure ~\ref{fig:category} shows the number of media organizations in each category in the dataset along with the percentage. Overall, we collected more than $2$ million Facebook posts. A Facebook post may contain a photo or a video or a link to an external source. In this paper, we limit ourselves to the link and video type posts only. This reduces the corpus size to $1.67$ million. For each post, we collect the headline (title of a video or headline of an article) and the status message. For a collection of $191,540$ link type posts, we also collected the bodies of the corresponding news articles. All these contents (headlines, messages, bodies) were used to train a domain specific word embeddings (details in Section ~\ref{sec-clickbaitDetection}). We also gather the Facebook reaction (Like, Love, Haha, Wow, Sad, Angry) statistics of each post. Table ~\ref{table:media_distribution} shows distribution of the corpus.

\begin{figure}[h]
	\centering
    \includegraphics[width=\linewidth]{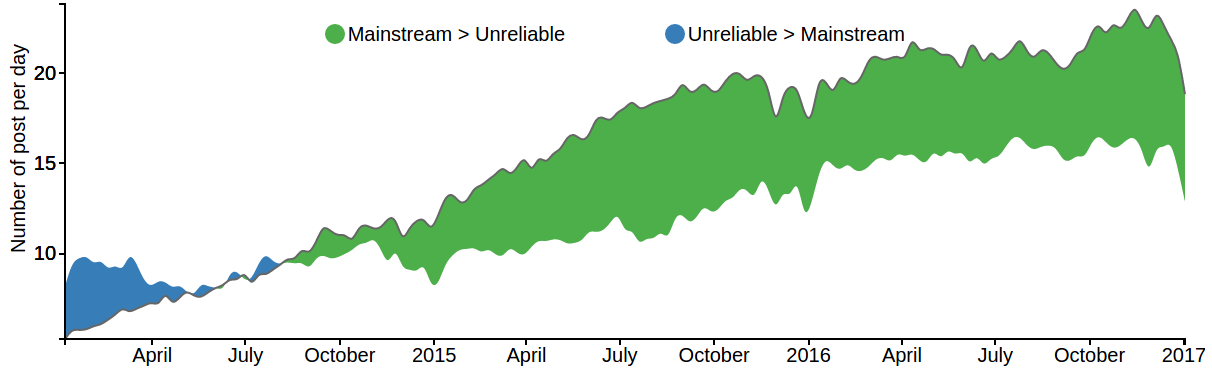}
    \caption{This figure shows the difference between the number of posts per day from an average mainstream (print, broadcast) media and the same from an average unreliable media during January $1^{st}$, 2014 -- December $31^{st}$, 2016. The green areas indicate that during these time periods, on average, a mainstream media posted more Facebook contents per day than an unreliable media. The blue areas indicate the opposite. General observation is, media organizations are sharing contents in the Facebook more actively now than they did earlier.}
    \label{fig:timeline}
\end{figure}

%The list of clickbait media is continuously curated by a group of media experts and researchers.~\footnote{www.niemanlab.org/2016/11/the-fake-news-wars-go-viral-with-melissas-list}
%We collect headlines and bodies of the articles referenced in the link type posts using the Python Newspaper package ~\footnote{https://pypi.python.org/pypi/newspaper}. We also collect titles of the videos from video type posts. In addition to the headline, body, and title, we also collect the accompanying Facebook status message  of the corresponding post.

\begin{table}[t]
\centering
\caption{Distribution of the \protect\dataset{2}}
\label{table:media_distribution}
\begin{tabular}{l|l|rr|r}
\toprule
				Media       & Category     & Link    & Video & Total   \\ \midrule
\multirow{2}{*}{Mainstream} & Broadcast    & 324028  & 32924 & 356952  \\
                            & Print        & 516713  & 14129 & 530842  \\ \midrule
\multirow{4}{*}{Unreliable} & Clickbait    & 371834  & 4099  & 375933  \\
                            & Conspiracy   & 309122  & 5841  & 314963  \\
                            & Junk Science & 51923   & 649   & 52572  \\
                            & Satire       & 41046   & 151   & 41197   \\ \midrule
\multicolumn{2}{c|}{Total}                 & 1614666 & 57793 & 1672459 \\ \bottomrule
\end{tabular}
\end{table}

\begin{figure}[h]
	\centering
    \includegraphics[width=0.7\linewidth]{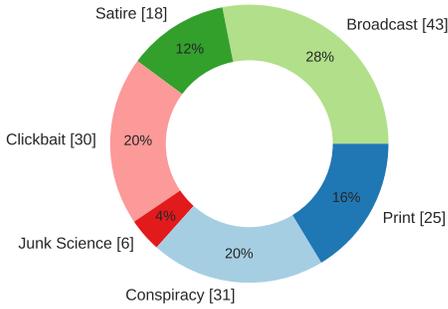}
    \caption{Category distribution of the \protect\dataset{2}}
    \label{fig:category}
\end{figure}
\section{Clickbait Detection}
\label{sec-clickbaitDetection}
The key purpose of this study is to systematically quantify the extents to which traditional print and broadcast media as well as ``alternative'' media -- often portrayed as unreliable -- use clickbait properties in contents published on the web.  The first step towards that goal is to identify clickbait and non-clickbait headlines.
\subsection{Problem Definition}
We define the clickbait identification task as a supervised binary classification problem where the set of classes, $\mathcal{C} =  \{clickbait, non\_clickbait\}$. Formally, given $\mathcal{X}$, a set of all sentences, and a training set $\mathcal{S}$ of labeled sentences  $\langle s, c\rangle$, where  $\langle s, c\rangle \in \mathcal{X} \times \mathcal{C}$, we want to learn a function $\gamma$ such that $\gamma : \mathcal{X} \rightarrow \mathcal{C}$, in other words, it maps sentences to $\{clickbait, non\_clickbait\}$. In the following sections, we describe modeling of the problem and compare performances of multiple learning techniques. 

% We want to classify each headline from the \dataset{1} dataset \((H)\) into two categories, named Clickbait (CB) and Non-Clickbait (NCB). So, our problem requires for each headline, \(h \in H, \) where \(H = \{h_1, h_2,....h_n\}\), we need to \(h := c_j, \) where \(c_j \in (CB, NCB)\)
\label{subsec-problemDefinition}

\subsection{Problem Modeling}
\label{subsec-problemModeling}
In text classification, a traditional approach is to use \emph{bag-of-words} (BOW) model to transform text into feature vectors before applying learning algorithms. ~\cite{chakraborty2016stop} followed this approach and used BOW model along with a collection of hand-crafted rules to prepare the feature set.
%However, a well-known limitation of BOW model is its invariance to order of words in a sentence.
However, inspired by the recent success of deep learning methods in text classification, we use distributed subword embeddings as features instead of applying BOW model. Specifically, we use an extension of the continuous \emph{skip-gram} model~\cite{mikolov2013distributed}, which takes into account subword (substring of a word) information ~\cite{bojanowski2016enriching}. We call this model as \clkbtwv. Below, we explain how \clkbtwv is used to generate word embeddings.

\subsubsection{\clkbtwv}
Given a large corpus $\mathcal{W}$, represented as a sequence of words, $\mathcal{W} = w_1, \dots, w_T$, the objective of the skip-gram model is to maximize the log-likelihood

\begin{equation}
\sum_{t=1}^T \sum_{c \in \mathcal{C}_t} \log p(w_c | w_t)
\label{eq:skipgram}
\end{equation}

where the context $\mathcal{C}_t$ is the set of indices of words surrounding $w_t$. In other words, given a word $w_t$, the model wants to maximize the correct prediction of its context $w_c$. The probability of observing a context word $w_c$ given $w_t$ is parametrized using the word vectors. The output of the model is an embedding for each word which captures semantic and contextual information of the word. \clkbtwv works in a slightly different way. Rather than treating each word as a unit, it breaks down words into subwords and wants to correctly predict the context subwords of a given subword. This extension allows sharing the representations across words, thus allowing to learn reliable representation for rare words. Consider the following example.

\begin{figure}[h]
	\centering
    \includegraphics[width=0.73\linewidth]{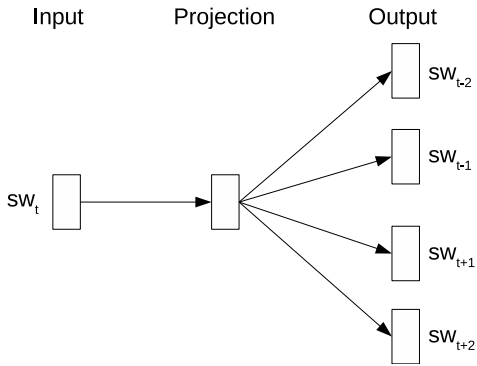}
    \caption{The \protect\clkbtwv model architecture. The training objective is to learn subword vector representations
that are good at predicting the nearby subwords.}
    \label{fig:subword_skipgram}
\end{figure}

\begin{example}
\emph{``the quick brown fox jumped over the lazy dog''}- take the word ``quick'' as an example. Assuming subword length as three, the subwords are- $\{qui, uic, ick\}$. \clkbtwv model learns to predict $qui$, $ick$ in the context given $uic$ as the input.
\end{example}

Figure ~\ref{fig:subword_skipgram} shows the architecture of the \clkbtwv model. Using neural network, the model learns the mapping between the output and the input. The weights to the hidden layer form the vector representations of the subwords. The embedding of a word is formed by the sum of the vector representations of its subwords. Formally, given a word $w$ and its set of subwords $\mathcal{SW}_w$, we can calculate the embedding of $w$ using the following equation-

\begin{equation}
\mathbf{u}_w = \sum_{sw \in \mathcal{SW}_w} \mathbf{v}_{sw}
\end{equation}

where $\mathbf{u}_w$ is the embedding of $w$ and $\mathbf{v}_{sw}$ is the vector representation of $sw$. Further details of the \clkbtwv model can be found in ~\cite{bojanowski2016enriching}.

\subsubsection{Pre-trained Vectors}
Note that \clkbtwv does not require $\mathcal{C}$ to learn the embeddings of words in corpus $\mathcal{W}$. It means that one can use the model on any large corpus of text to learn the word embeddings irrespective of whether the corpus is labeled or not. This technique of learning from large text corpus helps having richer word embeddings which capture a lot of semantic, conceptual and contextual information. We use the texts (headlines, messages, bodies) from \dataset{2} to learn word embeddings using this model. In Section~\ref{subsec-evaluation}, we present comparison between our pre-trained vectors and word vectors which were trained on about 100 billion words ~\cite{mikolov2013efficient} from the Google News dataset.

\subsubsection{Classification}
For a labeled sentence $\langle s, c\rangle$, we average the embeddings of words present in $s$ to form the hidden representation of $s$. These sentence representations are used to train a linear classifier. Specifically, we use the softmax function to compute the probability distribution over the classes in $\mathcal{C}$. ~\cite{joulin2016bag} describes the classification process in detail.

\subsection{Evaluation}
\label{subsec-evaluation}
We use the \dataset{1} dataset to evaluate our classification model. Section ~\ref{sec-dataset} provides the description of the dataset. We perform 10-fold cross-validation to evaluate various methods with respect to accuracy, precision, recall, f-measure, area under the ROC curve (ROC-AUC) and Cohen's $\kappa$. Table ~\ref{table:performance} shows performances of the methods. To avoid randomness effect, we perform each experiment $5$ times and present the average. There are in total seven methods. We categorize them based on the use of pre-trained vectors. Note that we report performances of Chakroborty et al. ~\cite{chakraborty2016stop} and Anand et al. ~\cite{anand2016we} in the table. We keep Anand et al. with the methods which use pre-trained vectors. Because Anand et al. used word embeddings trained on about 100 billion words from the Google News dataset using the Continuous Bag of Words architecture ~\cite{mikolov2013efficient}. Each word embedding has $300$ dimensions. Both of these works ~\cite{chakraborty2016stop,anand2016we} used a smaller and earlier version of the \dataset{1} dataset. Moreover, the training and test sets of the earlier dataset are not available. So, we could not compare our methods with them using the same test bed.

The \clkbtwv model, even without pre-trained vectors, significantly outperforms the BOW based Chakroborty et al. It achieves a f-measure score of $0.975$ ($2.5\%$ higher than Chakroborty et al.) and a $\kappa$ score of $0.952$. Powered with the pre-trained vectors, \clkbtwv performed even better. We used the same word embeddings provided by ~\cite{mikolov2013efficient} as well as our own \dataset{2}. Regarding the later, we experimented with three combinations- pre-trained vectors learned from the content headlines only, from headlines and messages, and from headline, bodies and messages. We set embedding size to $300$ dimensions while learning from these combinations. For the methods which were applied on the full \dataset{1} dataset, we highlight the top performance in each column. \clkbtwv along with pre-trained vectors from headlines, bodies and messages performed the best among all the variations. We realize that the differences of the measure values among the methods are small. However, we understand that making a small improvement while working above the $0.95$ range, is significant.

\dataset{2} has $477,236$ unique embeddings where Google News dataset provided $100$ billion embeddings. One interesting observation is, even though the size of our \dataset{2} is significantly smaller than the Google News dataset, it contributes more to the clickbait classification task. It can be rationalized as, the embeddings from \dataset{2} have more domain specific knowledge than the Google News dataset. We plan to extend this corpus with more Facebook posts and release it along with the pre-trained vectors for research purpose upon acceptance of the paper.

With this powerful clickbait classification model [\clkbtwv+(Headline+Body+Message)], we move forward and perform large scale study on the clickbait practice by a range of media on social network (Facebook). 
\begin{table*}[t]
\centering
% \small
\caption{Performance of the methods on the \protect\dataset{1} dataset}
% \resizebox{\columnwidth}{!}{%
\begin{threeparttable}
\begin{tabular}{l|l|rrrrrr}
\toprule
\multicolumn{2}{c}{Method}       	&	Precision	&	Recall	&	F-measure	&	Accuracy	&	Cohen's $\kappa$ & ROC-AUC\\
\midrule
\multirow{3}{*}{Without Pre-trained Vectors} & *Chakroborty et al.~\cite{chakraborty2016stop} & 0.95 & 0.90 & 0.93 & 0.93 & NA & 0.97 \\
& \clkbtwv		& 0.976 & 0.975	& 0.975	& 0.976 & 0.952	& 0.976 \\
\midrule
\multirow{4}{*}{With Pre-trained Vectors} & *Anand et al.~\cite{anand2016we} & 0.984 & 0.978 & 0.982 & 0.982 & NA & 0.998\\
& \clkbtwv + \googlewv		& 0.977          & 0.977		&	0.977		&	0.976		&	0.951 & 0.976 \\
& \clkbtwv + (Headline) & 0.981  & 0.981 & 0.981 & 0.981 & 0.962 & 0.981 \\
& \clkbtwv + (Headline + Message) & 0.982 & 0.982 & 0.982 &0.982 & 0.964 & 0.982 \\
& \clkbtwv + (Headline + Body + Message) & \textbf{0.983}   & \textbf{0.983} & \textbf{0.983} & \textbf{0.983} & \textbf{0.965} & \textbf{0.983} \\
\bottomrule
\end{tabular}%
% }
\begin{tablenotes}\footnotesize
\item[*] Their experiments were performed on a smaller and earlier version of the \dataset{1} dataset.
\end{tablenotes}
\end{threeparttable}
\label{table:performance}
\end{table*}
\section{Practice of using clickbait in Social Network}
\label{sec-clickbaitPractice}
We analyze the clickbait practice in Facebook using the \dataset{2} from three perspectives.

\begin{table}[!h]
\centering
\caption{Amount of clickbaits in various media}
\label{tab:clickbait}
\resizebox{\columnwidth}{!}{%
\begin{tabular}{l|l|rrr}
\toprule
Media                       & Category     & Clickbait & Non-clickbait & Clickbait (\%) \\ \midrule
\multirow{3}{*}{Mainstream} & Broadcast    & 169752    & 187200        & 47.56          \\
                            & Print        & 128022    & 402820        & 24.12          \\ \midrule
\multirow{3}{*}{Unreliable} & Clickbait    & 172271    & 203662        & 45.82          \\
                            & Conspiracy   & 90389     & 224574        & 28.7           \\
                            & Junk Science & 23637     & 28935         & 44.96          \\
                            & Satire       & 21798     & 19399         & 52.91          \\
\bottomrule
\end{tabular}%
}
\end{table}

% \reminder{Revise this section as we have new results now.}
\subsection{Quantitative Analysis}
\begin{figure}[h]
    \centering
    \begin{minipage}{0.45\linewidth}
        \centering
        \includegraphics[width=\linewidth, height=\linewidth]{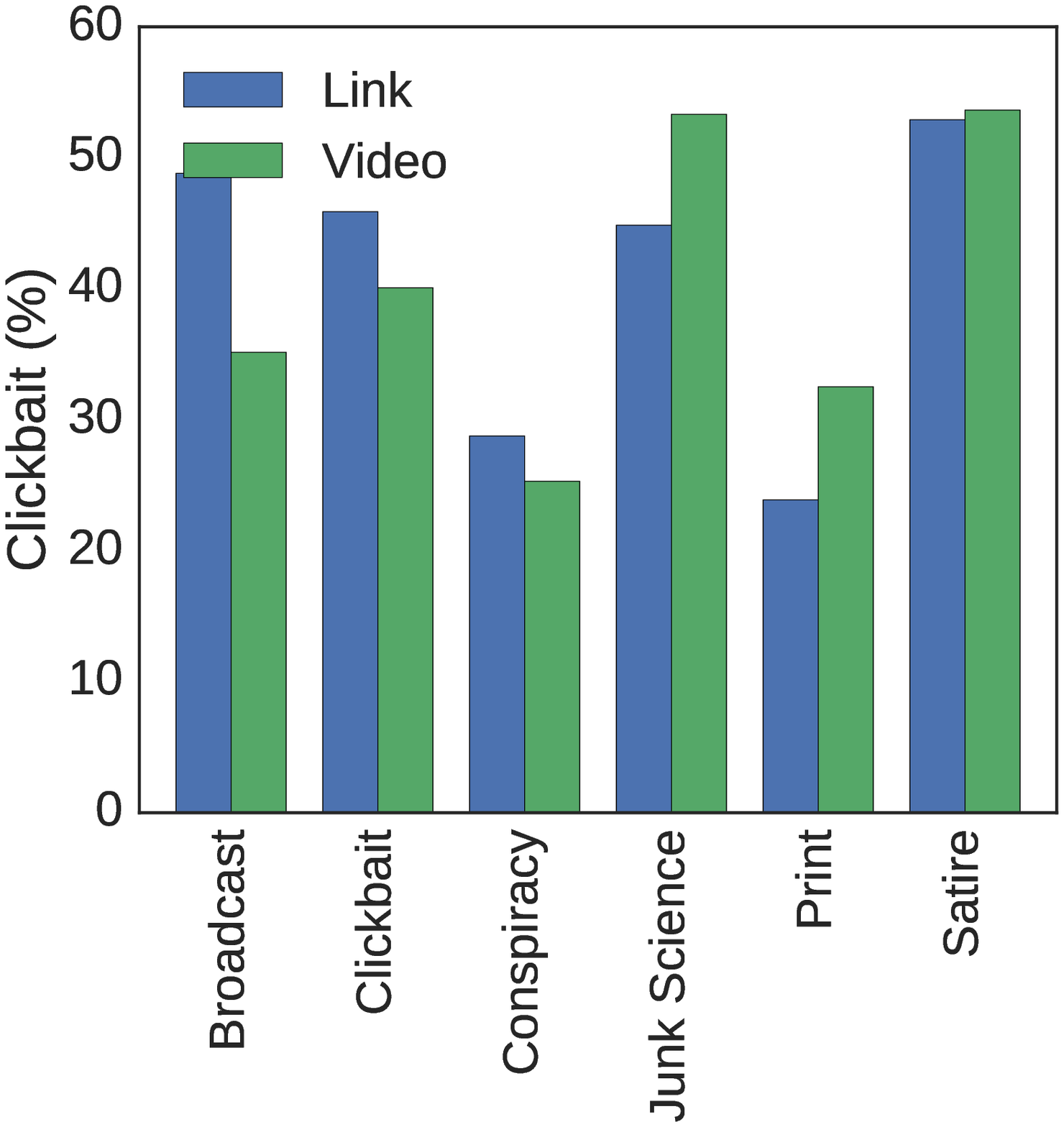}
        \caption{Percentage of clickbaits in link and video headlines.}
        \label{fig:link_video}
    \end{minipage}%
    \hspace{2mm}
    \begin{minipage}{0.45\linewidth}
        \centering
        \includegraphics[width=\linewidth, height=\linewidth]{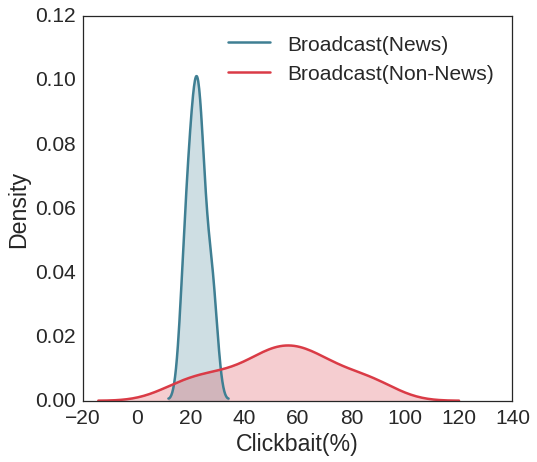}
        \caption{Broadcast (News) vs. Broadcast (Non-news).}
        \label{fig:broadcast}
    \end{minipage}
\end{figure}

To understand the extent of clickbait practice by different media and their categories, we applied the clickbait detection model on their contents; particularly on the headline/title of the link/video type posts. From now onward, we will use the term \emph{headline} to denote both the headline of a link content (article) and the title of a video content. Table ~\ref{tab:clickbait} shows amounts of clickbaits and non-clickbaits in the headlines of mainstream and unreliable media. Out of $887,794$ posts by mainstream media, $297,774$ $(33.54\%)$ have clickbait headlines. In unreliable media, the ratio is $39.26\%$ ($308,095$ clickbait headlines out of $784,665$). Based on these statistics, the percentage appears to be surprisingly high for the mainstream. We zoom into the categories of these two media to analyze the primary proponents of the clickbait practice. We find that between the two categories of mainstream media, broadcast uses clickbait $47.56\%$ of the times whereas print only uses $24.12\%$. We further zoom in to understand the high percentage in the broadcast category. The \dataset{2} has $43$ broadcast media. We manually categorize them into news oriented broadcast media (e.g. \emph{CNN}, \emph{NBC}, etc.) and non-news (lifestyle, entertainment, sports, etc.) broadcast media (e.g. \emph{HGTV}, \emph{E!}, etc.). There are $6$ news oriented broadcast media and $37$ non-news broadcast media. We find that the ratio of clickbait and non-clickbait is $61.64\%$ in non-news type broadcast media whereas it is only $22.32\%$ (close to print media) in news oriented media. Figure ~\ref{fig:broadcast} shows kernel density estimation of the clickbait percentage both for news and non-news broadcast media. It clearly shows the difference in clickbait practice in these two sub-categories. Most of the news type broadcast media has about $25\%$ clickbait contents. On the other hand, the percentage of clickbait for non-news type broadcast media has a wider range with peak at about $60\%$. In case of unreliable media, unsurprisingly all the categories have high percentage of clickbaits in their headlines. In Figure ~\ref{fig:link_video}, we show the percentage of clickbait in video and link type posts for each of the media categories. Satire is leading in both link and video type posts. Print and conspiracy media have the lowest clickbait practice among all the media categories in link and video type posts, respectively. Table ~\ref{tab:clickbait_practitioners} shows the top-$5$ clickbait proponents in each media category.

\begin{figure}[h]
	\centering
    \includegraphics[width=\linewidth]{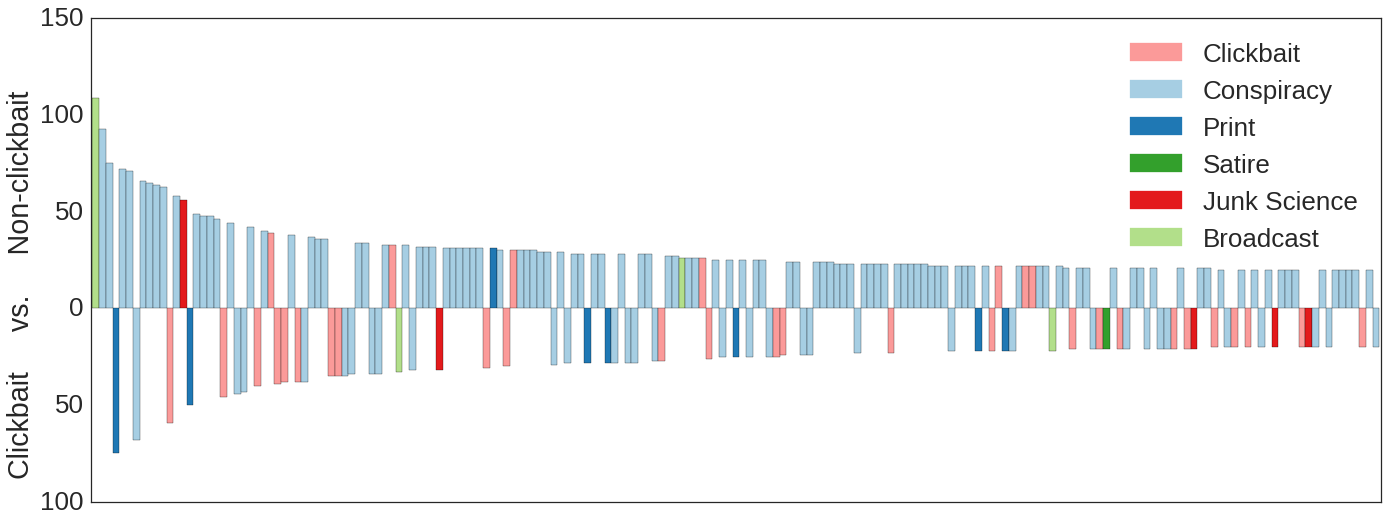}
    \caption{Frequency of link re-post by different media.}
    \label{fig:post_frequency}
\end{figure}

\subsection{Qualitative}
\textbf{Topic distribution}: To understand the topics in the clickbait and non-clickbait contents, we applied topic modeling on all the headlines of each category. One concern about applying the traditional topic modeling algorithms (e.g. Latent Dirichlet Allocation, Latent Semantic Analysis) on our corpus is, they focus on document-level word co-occurrence patterns to discover the topics of a document. So, they may struggle with the high word co-occurrence patterns sparsity which becomes a dominant factor in case of shorter context. That is why we use Biterm Topic Modeling (BTM) ~\cite{yan2013biterm} which generates the topics by directly modeling the aggregated word co-occurrence patterns of a short document.

\begin{table*}[t]
\centering
\caption{Topic model of Clickbait and Non-clickbait headlines in different media}
\label{tab:topic}
\resizebox{\textwidth}{!}{%
\begin{tabular}{l|l|l}
\toprule
Media                  & Clickbait & Non-Clickbait \\ \midrule
\multirow{5}{*}{Print}      & $T_1$: best, thing, day, new, 2015, cleveland, la, 2016, know, year            & $T_1$: new, san, la, jose, police, county, vega, get, bay, school                        \\ 
                                    & $T_2$: trump, woman, donald, new, get, say, make, people, thing, know          & $T_2$: police, man, cleveland, new, killed, woman, la, shooting, shot, get               \\ 
                                    & $T_3$: trump, new, get, woman, donald, make, star, say, man, chicago           & $T_3$: news, trump, new, man, say, york, woman, hawaii, police, killed                   \\ 
                                    & $T_4$: new, best, thing, year, get, kid, day, woman, make, trump               & $T_4$: trump, new, u, clinton, say, state, win, donald, take, world                      \\ 
                                    & $T_5$: boston, trump, donald, new, say, make, clinton, woman, get, 2016        & $T_5$: boston, new, say, trump, sox, chronicle, win, red, get, state                     \\ \midrule
\multirow{5}{*}{Broadcast}  & $T_1$: new, movie, star, make, swift, time, video, best, get, like             & $T_1$: police, man, new, found, woman, killed, arrested, say, shooting, death            \\ 
                                    & $T_2$: new, get, baby, kardashian, jenner, star, first, make, love, say        & $T_2$: trump, clinton, say, new, obama, u, gop, news, campaign, hillary,                 \\ 
                                    & $T_3$: woman, episode, new, trump, man, black, get, video, full, girl          & $T_3$: new, u, say, police, found, killed, dead, nbc, year, dy                           \\ 
                                    & $T_4$: trump, history, know, thing, donald, clinton, get, make, best, say      & $T_4$: win, new, say, game, first, get, team, player, take, back                         \\ 
                                    & $T_5$: day, photo, national, way, best, like, food, dog, thing, geographic     & $T_5$: national, geographic, photo, new, shark, day, classic, fs1undisputed, home, found \\ \midrule
\multirow{5}{*}{Unreliable} & $T_1$: trump, hillary, donald, clinton, obama, get, make, say, one, news       & $T_1$: obama, eagle, muslim, police, say, gun, u, cop, man, patriot                      \\ 
                                    & $T_2$: video, people, american, black, obama, muslim, u, america, cop, white   & $T_2$: trump, hillary, clinton, obama, new, say, campaign, news, donald, republican      \\ 
                                    & $T_3$: chick, trump, eagle, right, woman, hillary, say, get, people, make      & $T_3$: u, obama, video, war, isi, new, military, american, world, muslim                 \\ 
                                    & $T_4$: man, people, thing, woman, make, year, like, get, way, new              & $T_4$: new, truth, obama, say, u, republican, police, broadcast, man, american           \\ 
                                    & $T_5$: day, reunionfather, human, food, way, health, thing, reason, life, make & $T_5$: human, cancer, health, new, vaccine, u, study, food, found, world                          \\ \bottomrule
\end{tabular}%
}
\end{table*}

Table ~\ref{tab:topic} shows $5$ topics in clickbait and non-clickbait contents for each media category. Each topic is represented by a set of $10$ words. The words are ordered by their significance in the corresponding topic. The modeling indicates that clickbait headlines in print and broadcast media vary in tones and subject matters from their non-clickbait headlines to a great extent. Clickbait headlines in these media represent more personalized, sensationalized and entertaining topics, while non-clickbait headlines highlight topics of collective problems such as public policies and civic affairs. But this variation is not much evident in unreliable media that use clickbait headlines indiscriminately across all topics.

The model highlights some differences in clickbait topics between print and broadcast media. Most clickbait topics in print media, four out of five, are about U.S. President Donald Trump’s views on women. Each of these four topics include all of these four words: Trump, woman, make, new. A manual search shows that print news media often used clickbait techniques (e.g., question based headline) in stories about Trump and women. For instance, ``Did Donald Trump really say those things?'' was the headline of a Washington Post article dated July 25, 2016. The headline of a New York Times story from May 14, 2016, reads; ``Crossing the Line: How Donald Trump Behaved With Women in Private.''

Most clickbait topics in broadcast media are about entertainment (e.g., Taylor Swift’s new music video; Kardashian’s new baby) and lifestyle (e.g., food and health). Two topics appeared to touch Donald Trump and his opponent Hillary Clinton. Clickbait topics in unreliable media, however, range from politics to lifestyle. At least three topics appeared to be about politics in which key words include, Trump, Hillary, Obama, Muslim, Cop, and Woman. One topic is about food and health while another is unclear.

Non-clickbait topics remain similar across all three media types, which primarily focus on law and order, and U.S. presidential election campaign. Twelve out of 15 topics -- all five in print, three in broadcast, and four in unreliable -- are about these two areas. One broadcast topic appears to be about sports and one is unclear. One unreliable topic is about food and health.

%-Can we measure the distance between a headline and the actual content? Such a measure can help detect whether a headline is misleading or not. One idea is to calculate the average cosine similarity between the headline and the sentences in the first paragraph. The basis is, the first paragraph most likely [reference] will summarize the whole article. The lower the similarity, the more clickbaity the headline is.\\

\begin{figure}[h]
	\centering
    \includegraphics[width=\linewidth]{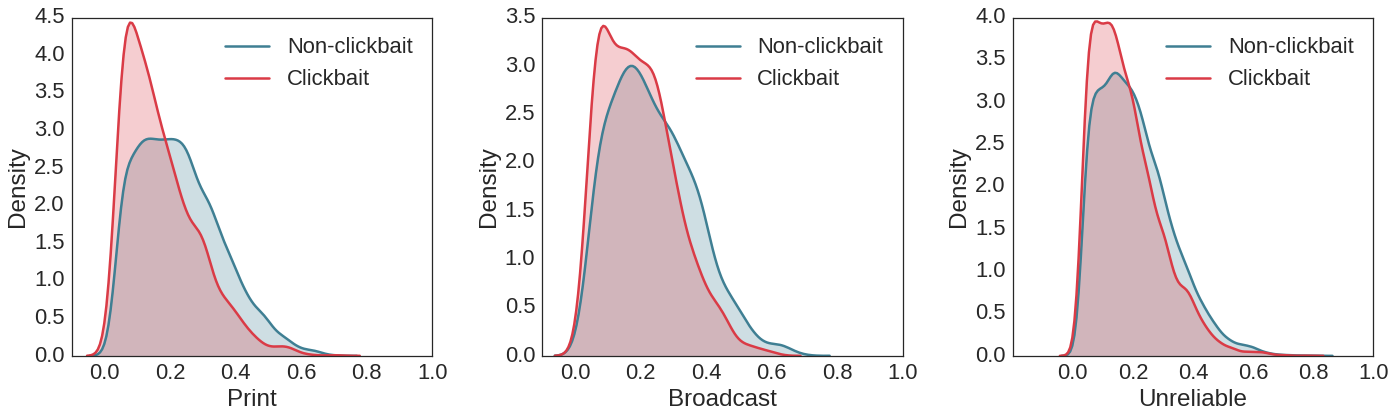}
    \caption{Headline-Body similarity in clickbait and non-clickbait contents.}
    \label{fig:similarity}
\end{figure}

\textbf{Headline-Body similarity}: One limitation of \clkbtwv is, it only considers the headline to determine whether it is a clickbait or not. The body of the news, is not considered as a factor in defining the headline. An attractive headline can be highly relevant to the content/body of a news or it can be very loosely related to the news. Our model is not capable of making the distinction. A metric is required to measure the similarity between the headline and the content to determine if the headline fairly represents the content. In future, we want to systematically incorporate the headline-body similarity in defining the clickbaitiness. Nonetheless, here we measure how similar the clickbait and non-clickbait headlines are to the corresponding bodies using a simple approach. We assume that the first para of an article represents the summary of the whole news ~\cite{izard1994fundamentals} and use cosine similarity to measure the similarity between the headline and the sentences in the first para. We use bag-of-words  model to transform the sentences into vectors before applying cosine similarity. In future, we plan to use our word embeddings to create the vectors instead. Figure ~\ref{fig:similarity} shows the kernel density estimation of the headline-body similarity in clickbait and non-clickbait contents posted by different media. One observation is, in print media non-clickbait headlines are closer to their summary than clickbait headlines. In broadcast media, the difference is less clear and in unreliable media the difference is almost absent.

\begin{table}[t]
\centering
\caption{Top-5 clickbait proponents in each media}
\label{tab:clickbait_practitioners}
\resizebox{\columnwidth}{!}{%
\begin{tabular}{l|l|rrr}
\toprule
Media                       & Name                & Clickbait & Non-clickbait & Clickbait (\%) \\ \midrule
\multirow{5}{*}{Overall}    & VH1                 & 13760     & 1339          & 91.13         \\
                            & AmplifyingGlass     & 692       & 71            & 90.69         \\
                            & MTV                 & 42313     & 4492          & 90.4          \\
                            & ClickHole           & 8250      & 930           & 89.87         \\
                            & Reductress          & 3984      & 484           & 89.17         \\ \midrule
\multirow{5}{*}{Broadcast}  & VH1                 & 13760     & 1339          & 91.13         \\
                            & MTV                 & 42313     & 4492          & 90.4          \\
                            & Bravo TV            & 8263      & 1242          & 86.93         \\
                            & Food Network        & 2990      & 492           & 85.87         \\
                            & OWN                 & 474       & 118           & 80.07         \\ \midrule
\multirow{5}{*}{Print}      & Washington Post     & 13905     & 15158         & 47.84         \\
                            & New York Post       & 11977     & 13910         & 46.27         \\
                            & Dallas Morning News & 3982      & 8232          & 32.6          \\
                            & USA Today           & 8538      & 20282         & 29.63         \\
                            & Houston Chronicle   & 8481      & 21618         & 28.18         \\ \midrule
\multirow{5}{*}{Unreliable} & AmplifyingGlass     & 692       & 71            & 90.69         \\
                            & ClickHole           & 8250      & 930           & 89.87         \\
                            & Reductress          & 3984      & 484           & 89.17         \\
                            & Food Babe           & 2387      & 638           & 78.91         \\
                            & Chicks on the Right & 14185     & 4977          & 74.03         \\
\bottomrule
\end{tabular}%
}
\end{table}

\begin{table}[b]
\centering
\caption{Presence of clickbait in the status}
\label{tab:status}
\resizebox{\columnwidth}{!}{%
\begin{tabular}{l|l|rrr}
\toprule
Media						& Category         & Clickbait Status & Non-clickbait Link & Clickbait Status (\%)      \\
\midrule
\multirow{2}{*}{Mainstream} &  Broadcast    & 84192  & 176177 & 32.34 \\
& Print        & 164669 & 379504 & 30.26 \\ \midrule
\multirow{2}{*}{Unreliable} &  Clickbait    & 91747  & 157886 & 36.75 \\
& Conspiracy   & 46851  & 190477 & 19.74 \\
& Junk Science & 12764  & 28349  & 31.05 \\
& Satire       & 7425   & 14453  & 33.94 \\
\bottomrule                            
\end{tabular}%
}
\end{table}

\subsection{Impact}
To measure the reachability and user engagement of clickbait and non-clickbait contents, we use Facebook reactions, comments and shares as metrices. Figure 8 shows number of comments, shares and reactions (summation of like, haha, wow, sad, angry, happy, love) of an average clickbait and non-clickbait post in each media category. Blue areas indicate that on average, a clickbait post (link or video) receives more attention (reactions/shares/comments) than a non-clickbait post. Green areas indicate the opposite. Clickbait contents receive more attention and reach to more users in general. One exception is the broadcast media.

We also analyze how often a news article is re-posted in Facebook. Figure ~\ref{fig:post_frequency} shows number of times a link is re-posted by a media. Each bar represents a news link. The height indicates how many times this link was posted in Facebook by the colored media category. We only consider the links which were re-posted at least 20 times. Compare to others, conspiracy media organizations repeat the same link more. This is observed both for clickbait and non-clickbait. Clickbait media seem to repeatedly posting same clickbait links more than others.

Other than headlines, the media organizations also practice using clickbait in the Facebook status message itself. Table ~\ref{tab:status} shows use of clickbait status for non-clickbait articles by different media. A general observation is, the practice is there to allure the readers by giving clickbaity message posts even for non-clickbaity news contents. Unsurprisingly, the clickbait media category is leading in this practice.

\begin{figure*}[!h]
    \centering
    \begin{minipage}{0.33\linewidth}
        \centering
        \includegraphics[width=\linewidth ]{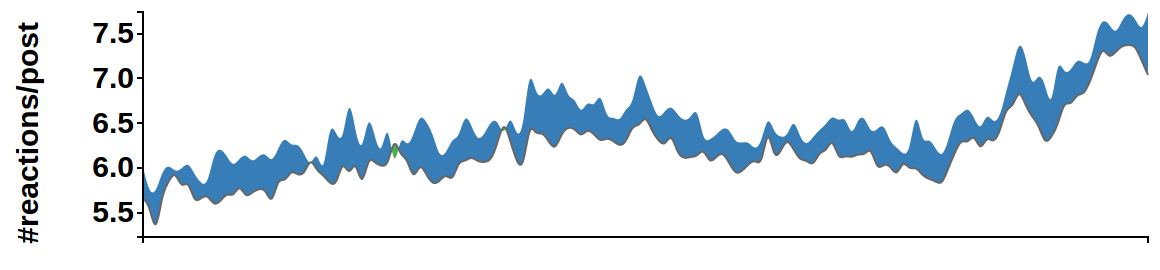}
%         \caption{Presence of clickbaits in links and videos}
        \label{fig:pr}
    \end{minipage}%
%     \hspace{2mm}
    \begin{minipage}{0.33\linewidth}
        \centering
        \includegraphics[width=\linewidth ]{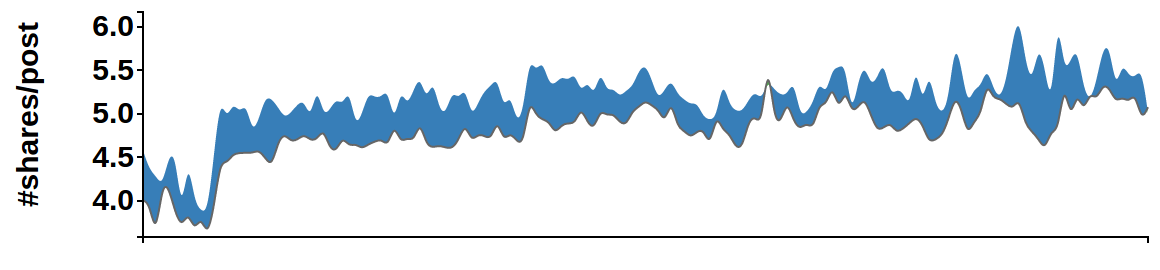}
%         \caption{Broadcast (News) vs. Broadcast (Non-news)}
        \label{fig:ps}
    \end{minipage}%
%     \hspace{2mm}
    \begin{minipage}{0.33\linewidth}
        \centering
        \includegraphics[width=\linewidth ]{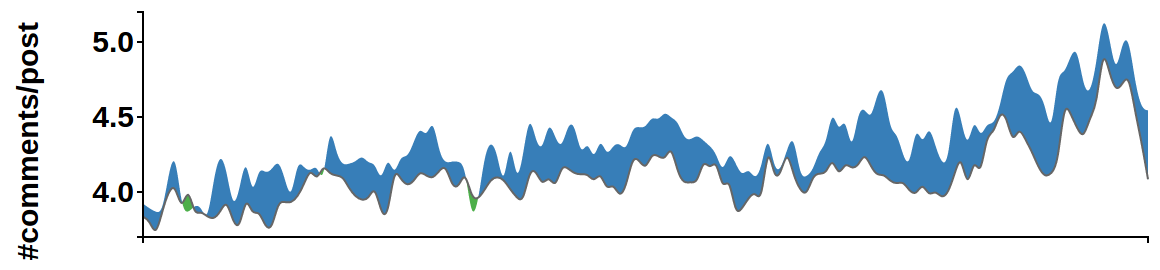}
%         \caption{Broadcast (News) vs. Broadcast (Non-news)}
        \label{fig:pc}
    \end{minipage}
    
    \begin{minipage}{0.33\linewidth}
        \centering
        \includegraphics[width=\linewidth ]{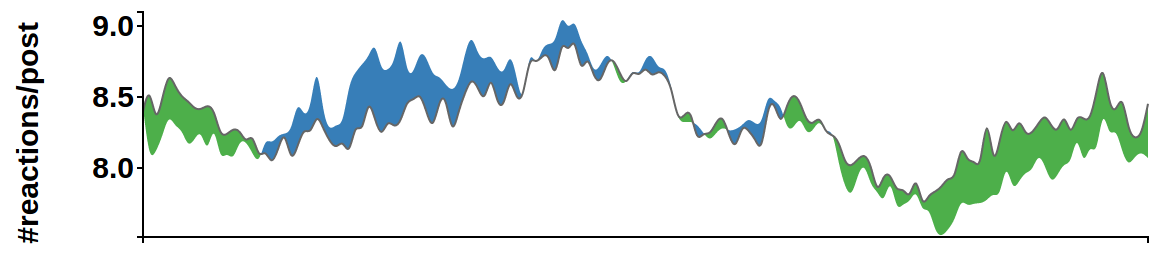}
%         \caption{Presence of clickbaits in links and videos}
        \label{fig:br}
    \end{minipage}%
%     \hspace{2mm}
    \begin{minipage}{0.33\linewidth}
        \centering
        \includegraphics[width=\linewidth]{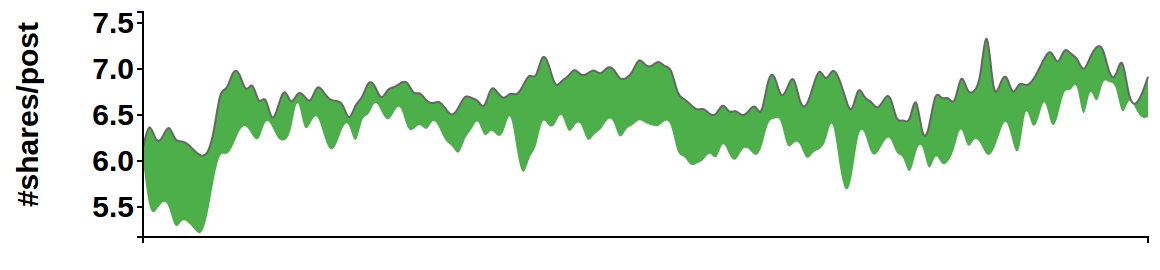}
%         \caption{Broadcast (News) vs. Broadcast (Non-news)}
        \label{fig:bs}
    \end{minipage}%
%     \hspace{2mm}
    \begin{minipage}{0.33\linewidth}
        \centering
        \includegraphics[width=\linewidth ]{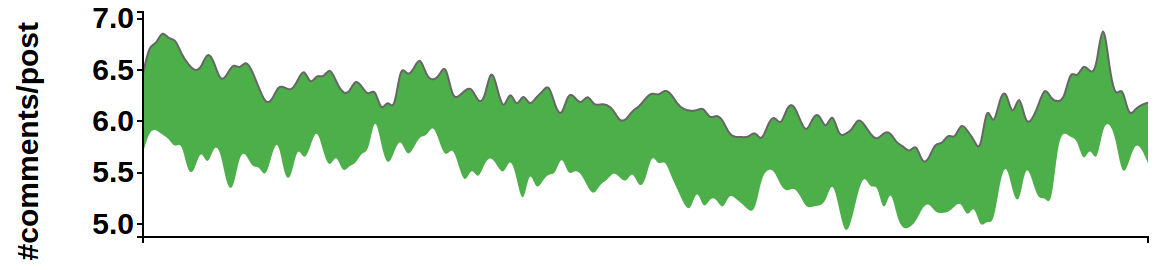}
%         \caption{Broadcast (News) vs. Broadcast (Non-news)}
        \label{fig:bc}
    \end{minipage}
    
    \begin{minipage}{0.33\linewidth}
        \centering
        \includegraphics[width=\linewidth ]{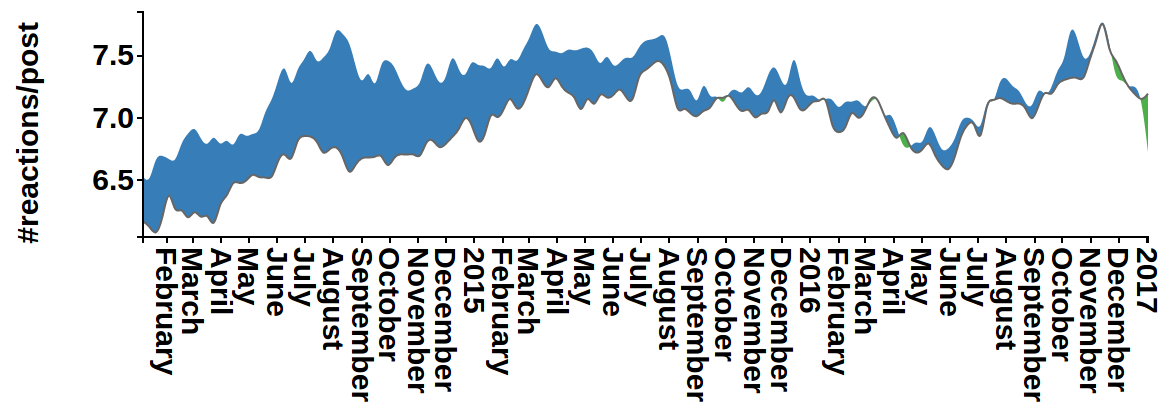}
%         \caption{Presence of clickbaits in links and videos}
        \label{fig:ur}
    \end{minipage}%
%     \hspace{2mm}
    \begin{minipage}{0.33\linewidth}
        \centering
        \includegraphics[width=\linewidth]{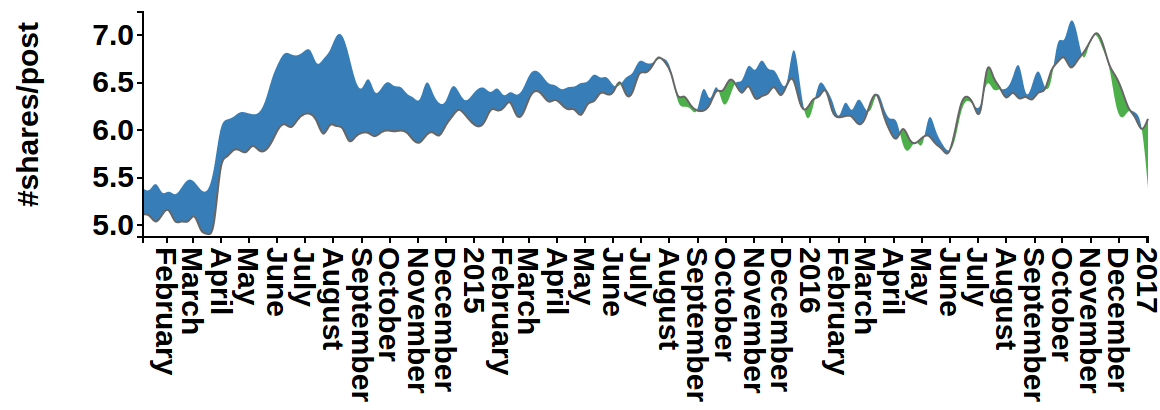}
%         \caption{Broadcast (News) vs. Broadcast (Non-news)}
        \label{fig:us}
    \end{minipage}%
%     \hspace{2mm}
    \begin{minipage}{0.33\linewidth}
        \centering
        \includegraphics[width=\linewidth ]{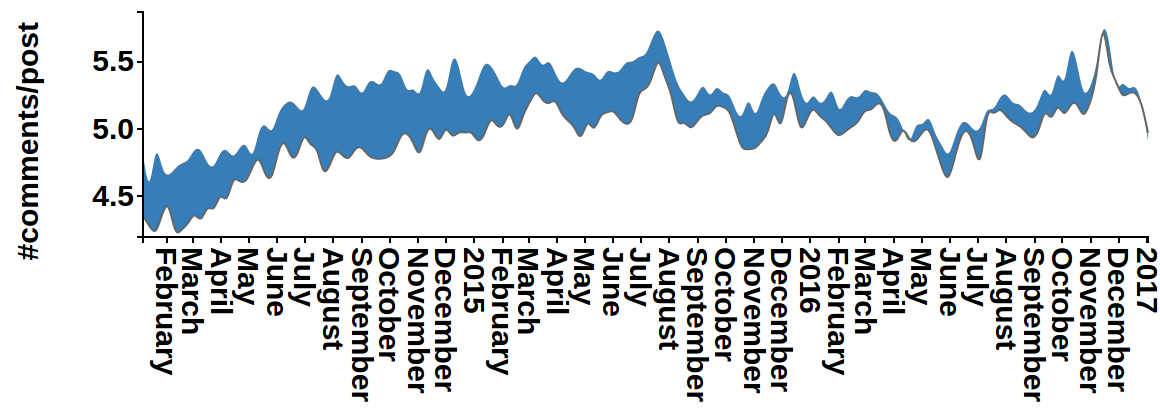}
%         \caption{Broadcast (News) vs. Broadcast (Non-news)}
        \label{fig:uc}
    \end{minipage}
    \label{fig:facebook}
    \caption{Top: Print media, Middle: Broadcast media, Bottom: Unreliable media. Blue areas indicate that on average, a clickbait post (link or video) receives more attention (reactions/shares/comments) than a non-clickbait post. Green areas indicate the opposite. }
\end{figure*}
\section{Related Work}
\label{sec-relatedWork}
Even though \emph{clickbait} is a relatively nascent term, its traces can be found in several journalistic concepts such as \emph{tabloidization} and \emph{content trivialization}. The linguistic techniques and presentation styles, employed typically in clickbait headlines and articles, derived from the tabloid press that baits readers with sensational language and appealing topics such as celebrity gossip, humor, fear and sex~\cite{palau2016reference}. Clickbait articles are also similar to tabloid press articles in terms of story focus, which puts emphasis on the entertaining elements of an event rather than  the informative elements. The Internet and especially the social media have made it easier for the clickbait practitioners to create, publish in a larger scale and reach to a broader audience with a higher speed than before~\cite{ingram2016gigacom}. In the last several years, academicians and media studied this phenomenon from several perspectives.

\textbf{Clickbait-- Properties, Practice and Effects}: There have been a small number of studies--some conducted by academic researchers and others by media  firms--which examined correlations between headline attributes and degree of user engagement with content. Some media market analysts and commentators ~\cite{filloux2016clickbait} discussed various aspects of this practice. However, no research has been found, which gauges the extents of clickbait practices by mainstream and alternative media outlets on the web. Nor have we found any study that examined if clickbait techniques help increase user engagement on social media.

A journalism professor~\cite{palau2016reference} manually examined content of four online sections of the Spanish newspaper \emph{El Pais} \footnote{http://elpais.com}, which apparently used clickbait features to capture attention. The corpus included only $151$ articles published in June, 2015. The articles in the corpus appeared to emphasize anecdotal aspects, or issues with little value, and curiosities. The study identified various linguistic techniques used in headlines of these articles such as orality markers and interaction (e.g., direct appeal to the reader), vocabulary and word games (e.g., informal language, generic or buzzwords), and morphosyntax (e.g., simple structures).

Researchers at the University of Texas's \emph{Engaging News Project} ~\cite{joshua2016ashley}  conducted an experiment on $2,057$ U.S. adults to examine their reactions to clickbait (e.g., question-based headlines) and traditional news headlines in political articles. They found that clickbait headlines led to more negative reactions among users than non-clickbait headlines. Interestingly, the same users were slightly more engaged with non-traditional media that tend to use clickbait techniques more often. This finding questions the conventional belief that user reactions may predict user engagement, and warrants large-scale investigations. 

\emph{Chartbeat}, an analytics firm that provides market intelligence to media organizations, tested $10,000$ headlines from over 100 websites for their effectiveness in engaging users with content ~\cite{chartbeat2015clickbait}. The study examined $12$ `common tropes' in headlines-- a majority of them are considered clickbait techniques -- and found that some of these tropes are more effective than others. Some media pundits interpreted the findings of this study as clickbaits being detrimental to traditional news brands.

\emph{HubSpot} and \emph{Outbrain}, two content marketing platforms that distribute clickbait contents across the web, examined millions of headlines to identify attributes that contribute to traffic growth, increased engagement, and conversion of readers into subscribers. The study suggested that clickbait techniques may increase temporary engagement ~\cite{outbrain2016hubspot}, but an article must deliver on its promises made in headline for users to return and convert.

\textbf{Automated Clickbait Detection}: \cite{potthast2016clickbait, chakraborty2016stop,anand2016we,abhishek2015clickbait} study automated detection of clickbait headlines using natural language processing and machine learning. \cite{abhishek2015clickbait} collects $10,000$ headlines from \emph{Buzzfeed}, \emph{Clickhole}, and \emph{The New York Times (NYT)} and uses Logistic Regression to create a supervised clickbait detection model. It assumes all \emph{Buzzfeed} and \emph{Clickhole} headlines as clickbait and all \emph{NYT} headlines as non-clickbait. We would like to argue that it makes the model susceptible to personal bias as it overlooks the fact that many \emph{Buzzfeed} contents are original, non-clickbaity and there are clickbait practice in \emph{NYT}~\cite{hurst2016clickbait}. Moreover, \emph{BuzzFeed}, and \emph{NYT} usually write on very different topics. The model might have been trained merely as a topic classifier. ~\cite{potthast2016clickbait} attempts to detect clickbaity Tweets in Twitter by using common words occurring in clickbaits, and by extracting some tweet specific features. ~\cite{chakraborty2016stop} uses a dataset of $15,000$ manually labeled headlines to train several supervised models for clickbait detection. These methods heavily depend on a rich set of hand-crafted features which take good amount of time to engineer and sometimes are specific to the domain (for example, tweet related features are specific to Twitter data and inapplicable to other domains). ~\cite{anand2016we} presents clickbait detection model which uses word embeddings and Recurrent Neural Network (RNN). These works consider the structure and semantic of a headline to determine whether it is a clickbait or not. However, one important aspect, the body of the news, is not considered as a factor in these works at all. We would like to argue that only the headline itself does not fully represent whether an article is a clickbait or not. If a headline represents the body fairly, it should not be considered as a clickbait. Consider the title as an example, \textit{`The Top 10 Mistakes Of Entrepreneurs'}\footnote{www.forbes.com/sites/roberthof/2016/02/23/guy-kawasaki-the-top-10-mistakes-of-entrepreneurs}. It is as clickbait of a headline as it can be. However, the body actually contains reasonably decent materials, which might be interesting to many people.

\textbf{Clickbait Generation} ~\cite{clickbait2015generator,cha2015generator,linkbait2015generator} present automated clickbait generation tools. ~\cite{clickbait2015generator} trains an RNN model using $2$ million headlines collected from \emph{Buzzfeed}, \emph{Gawker}, \emph{Jezebel}, \emph{Huffington Post} and \emph{Upworthy}. The model is then used to produce new clickbait headlines

%There needs to be a systematic way to measure the distance between the headline and the content of a news article to determine if the headline fairly represents the content or not. 

\section{Conclusion}
\label{sec-conclusion}
% \begin{itemize}
% \item We are not considering the context of the news articles. In future, we would like to leverage the contents as well as the headlines.
% \item We want to measure the distance between a headline and the corresponding content more systematically.
% \item Cluster similar posts belonging to a news and see which media uses clickbait to report that news and which doesn't.\\
% \end{itemize}
In this paper, we introduce a word-embedding based clickbait detection system which is built on our own collected corpus of news headlines and contents. We showed that our model performs better than the Google news dataset based embeddings. Our analysis also reveals how mainstream media are getting involved into clickbait practicing increasingly. Close scrutiny of the social media posts also reveals that broadcast type media has higher percentage of usage of clickbait practice than the print media and non-news type broadcast media mostly contributes to it. Our study also brings forth another fact of using higher percentage of clickbait practice by unreliable media which is quite obvious. Moreover, results from our topic modeling indicates that clickbait practice is prevalent in personalized and entertaining areas. In future, we want to incorporate the content of the news in defining the clickbaitiness of a headline. We believe, such a system would help social networking platforms to curb the problem of clickbait and provide a better using experience. %Finally, our analysis of the similarity between the news content and its headline paves the need of more systematic approach of exploring this.

\IEEEpeerreviewmaketitle

\bibliographystyle{IEEEtran}
\bibliography{references}

\begin{thebibliography}{10}
\providecommand{\url}[1]{#1}
\csname url@samestyle\endcsname
\providecommand{\newblock}{\relax}
\providecommand{\bibinfo}[2]{#2}
\providecommand{\BIBentrySTDinterwordspacing}{\spaceskip=0pt\relax}
\providecommand{\BIBentryALTinterwordstretchfactor}{4}
\providecommand{\BIBentryALTinterwordspacing}{\spaceskip=\fontdimen2\font plus
\BIBentryALTinterwordstretchfactor\fontdimen3\font minus
  \fontdimen4\font\relax}
\providecommand{\BIBforeignlanguage}[2]{{%
\expandafter\ifx\csname l@#1\endcsname\relax
\typeout{** WARNING: IEEEtran.bst: No hyphenation pattern has been}%
\typeout{** loaded for the language `#1'. Using the pattern for}%
\typeout{** the default language instead.}%
\else
\language=\csname l@#1\endcsname
\fi
#2}}
\providecommand{\BIBdecl}{\relax}
\BIBdecl

\bibitem{palau2016reference}
D.~Palau-Sampio, ``Reference press metamorphosis in the digital context:
  clickbait and tabloid strategies in elpais. com.'' \emph{Communication \&
  Society}, vol.~29, no.~2, 2016.

\bibitem{chakraborty2016stop}
A.~Chakraborty, B.~Paranjape, S.~Kakarla, and N.~Ganguly, ``Stop clickbait:
  Detecting and preventing clickbaits in online news media,'' in \emph{Advances
  in Social Networks Analysis and Mining (ASONAM), 2016 IEEE/ACM International
  Conference on}.\hskip 1em plus 0.5em minus 0.4em\relax IEEE, 2016, pp. 9--16.

\bibitem{mark2016adele}
M.~S. LUCKIE, ``Adele and the death of clickbait,''
  \url{http://www.niemanlab.org/2015/12/adele-and-the-death-of-clickbait/},
  2015.

\bibitem{chris2016clickbait}
C.~Sutcliffe, ``Can publishers step away from the brink of peak content?''
  \url{https://www.themediabriefing.com/article/can-publishers-step-away-from-the-brink-of-peak-content},
  2016.

\bibitem{joshua2016ashley}
J.~M.~Scacco and A.~Muddiman, ``Investigating the influence of ``clickbait''
  news headlines,''
  \url{https://engagingnewsproject.org/wp-content/uploads/2016/08/ENP-Investigating-the-Influence-of-Clickbait-News-Headlines.pdf},
  2016.

\bibitem{bojanowski2016enriching}
P.~Bojanowski, E.~Grave, A.~Joulin, and T.~Mikolov, ``Enriching word vectors
  with subword information,'' \emph{arXiv preprint arXiv:1607.04606}, 2016.

\bibitem{joulin2016bag}
A.~Joulin, E.~Grave, P.~Bojanowski, and T.~Mikolov, ``Bag of tricks for
  efficient text classification,'' \emph{arXiv preprint arXiv:1607.01759},
  2016.

\bibitem{mikolov2013distributed}
T.~Mikolov, I.~Sutskever, K.~Chen, G.~S. Corrado, and J.~Dean, ``Distributed
  representations of words and phrases and their compositionality,'' in
  \emph{Advances in neural information processing systems}, 2013, pp.
  3111--3119.

\bibitem{mikolov2013efficient}
T.~Mikolov, K.~Chen, G.~Corrado, and J.~Dean, ``Efficient estimation of word
  representations in vector space,'' \emph{arXiv preprint arXiv:1301.3781},
  2013.

\bibitem{anand2016we}
A.~Anand, T.~Chakraborty, and N.~Park, ``We used neural networks to detect
  clickbaits: You won't believe what happened next!'' \emph{arXiv preprint
  arXiv:1612.01340}, 2016.

\bibitem{nielson2016fcc}
J.~Eggerton, ``Fcc: Nielsen dmas still best definition of tv market,''
  Broadcasting \& Cable, 2016.

\bibitem{info2016list}
informationisbeautiful.net, ``Unreliable/fake news sites \& sources,''
  \url{https://docs.google.com/spreadsheets/d/1xDDmbr54qzzG8wUrRdxQl_C1dixJSIYqQUaXVZBqsJs},
  2016.

\bibitem{melissa2016list}
M.~Zimdars, ``My `fake news list' went viral. but made-up stories are only part
  of the problem,''
  \url{https://www.washingtonpost.com/posteverything/wp/2016/11/18/my-fake-news-list-went-viral-but-made-up-stories-are-only-part-of-the-problem},
  2016.

\bibitem{yan2013biterm}
X.~Yan, J.~Guo, Y.~Lan, and X.~Cheng, ``A biterm topic model for short texts,''
  in \emph{Proceedings of the 22nd international conference on World Wide
  Web}.\hskip 1em plus 0.5em minus 0.4em\relax ACM, 2013, pp. 1445--1456.

\bibitem{izard1994fundamentals}
R.~S. Izard, H.~M. Culbertson, and D.~A. Lambert, \emph{Fundamentals of news
  reporting}.\hskip 1em plus 0.5em minus 0.4em\relax Kendall/Hunt Publishing
  Company, 1994.

\bibitem{ingram2016gigacom}
M.~Ingram, ``The internet didn’t invent viral content or clickbait journalism
  -- there’s just more of it now, and it happens faster,''
  \url{https://gigaom.com/2014/04/01/the-internet-didnt-invent-viral-content-or-clickbait-journalism-theres-just-more-of-it-now-and-it-happens-faster},
  2014.

\bibitem{filloux2016clickbait}
F.~Filloux, ``Clickbait is devouring journalism but there are ways out,''
  \url{https://qz.com/648845/clickbait-is-devouring-journalism-but-there-are-ways-out/},
  2016.

\bibitem{chartbeat2015clickbait}
C.~Breaux, ``You'll never guess how chartbeat's data scientists came up with
  the single greatest headline,''
  \url{http://blog.chartbeat.com/2015/11/20/youll-never-guess-how-chartbeats-data-scientists-came-up-with-the-single-greatest-headline},
  2015.

\bibitem{outbrain2016hubspot}
Hubspot and Outbrain, ``Data driven strategies for writing effective titles \&
  headlines,''
  \url{http://cdn2.hubspot.net/hub/53/file-2505556912-pdf/Data_Driven\_Strategies\_For\_Writing\_Effective\_Titles\_and\_Headlines.pdf}.

\bibitem{potthast2016clickbait}
M.~Potthast, S.~K{\"o}psel, B.~Stein, and M.~Hagen, ``Clickbait detection,'' in
  \emph{European Conference on Information Retrieval}.\hskip 1em plus 0.5em
  minus 0.4em\relax Springer, 2016, pp. 810--817.

\bibitem{abhishek2015clickbait}
A.~Thakur, ``Identifying clickbaits using machine learning,''
  \url{https://www.linkedin.com/pulse/identifying-clickbaits-using-machine-learning-abhishek-thakur},
  2016.

\bibitem{hurst2016clickbait}
N.~Hurst, ``To clickbait or not to clickbait? an examination of clickbait
  headline effects on source credibility,'' Ph.D. dissertation, University of
  Missouri--Columbia, 2016.

\bibitem{clickbait2015generator}
L.~Eidnes, ``Auto-generating clickbait with recurrent neural networks,''
  \url{https://larseidnes.com/2015/10/13/auto-generating-clickbait-with-recurrent-neural-networks},
  2015.

\bibitem{cha2015generator}
C.~Cha, ``clickbait generator,'' \url{http://www.thisisreallyreal.com/}, 2016.

\bibitem{linkbait2015generator}
``Linkbait title generator,''
  \url{http://www.contentrow.com/tools/link-bait-title-generator}.

\end{thebibliography}

\end{document}